\documentclass[preprint,showpacs,preprintnumbers,amsmath,amssymb]{revtex4}
\usepackage[dvips]{epsfig}
\def\be{\begin{equation}}
\def\ee{\end{equation}}
\def\bdm{\begin{eqnarray}}
\def\edm{\end{eqnarray}}
\begin{document}
\preprint{Submitted to Physics of Plasmas}
\title{Analytical description of stochastic field-line wandering in magnetic turbulence}
\author{A. Shalchi \& I. Kourakis}
\affiliation{Institut f\"ur Theoretische Physik, Lehrstuhl IV:
Weltraum- und Astrophysik, Ruhr-Universit\"at Bochum, D-44780
Bochum, Germany}
\date{\today}
\begin{abstract}
A non-perturbative nonlinear statistical approach is presented to
describe turbulent magnetic systems embedded in a uniform mean
magnetic field. A general formula in the form of an ordinary
differential equation for magnetic field-line wandering (random
walk) is derived. By considering the solution of this equation for
different limits several new results are obtained. As an example,
it is demonstrated that the stochastic wandering of magnetic
field-lines in a two-component turbulence model leads to
superdiffusive transport, contrary to an existing diffusive
picture. The validity of quasilinear theory for field-line
wandering is discussed, with respect to different turbulence
geometry models, and previous diffusive results are shown to be
deduced in appropriate limits.
\end{abstract}
\pacs{47.27.tb, 96.50.Ci, 96.50.Bh} \maketitle
\section{Introduction}

Understanding turbulence occupies a central part of current
research efforts in space physics and astrophysics; see, e.g., in
Refs. \cite{MC,RS}. Charged particles in turbulent magnetized
environments perform a complex motion, which may be seen as the
sum of a deterministic trajectory (helical motion around the
magnetic field lines) and a random component, due to turbulence.
In an effort to elucidate transport mechanisms in turbulent
(collisionless) magnetized plasmas, some progress has been marked
by employing a statistical description of turbulence (e.g.
\cite{gol95, cho02, zho04}). The erratic component of charged
particle motion can physically be associated with the stochastic
wandering (random walk) of magnetic field-lines; see Refs. {\bf
\cite{Jo73, Skill74, Nar01, Matt03, Chan04, Maron04, kot00,
web06}}. It is nevertheless understood that a definite, widely
applicable analytical formulation of FLRW (field-line random walk)
is not yet available.

A quasi-linear approach for field-line random-walk formed the
model basis in early works; see ,e.g., Ref. \cite{jok66}. In that
description, the unperturbed field-lines are used to describe
field-line wandering by using a perturbation method. This approach
is believed to be correct in the limit of weak turbulence, where
turbulent fields are assumed to be much weaker than the (uniform)
mean field ($\delta B_i \ll B_0$). A non-perturbative statistical
description of field-line wandering was later suggested in Ref.
\cite{mat95}, relying on certain assumptions about the properties
of the field-lines (e.g. Gaussian statistics) in combination with
an explicit diffusive hypothesis for the field-line topology.

The main scope of this article is to address the problem of
field-line wandering analytically, in relation with general
turbulence models. The limits of the validity of quasilinear
theory (QLT) are also to be discussed. The standard statistical
description of field-line wandering is rigorously shown to lead to
an ordinary differential equation (ODE) for the mean square
deviation (MSD) of the field-lines. The ODE is solved in certain
cases and the results are compared with QLT and associated
methods. An interesting example is field-line random walk in
two-component turbulence, where a superdiffusive behavior of
field-line wandering is clearly found.
\section{An ODE for field-line wandering}
We shall consider a collisionless magnetized plasma system which
is embedded in a uniform mean field ($\vec{B}_0 = B_0 \vec{e}_z$)
in addition to a turbulent magnetic field component in the
transverse direction ($\delta B_z = 0$). The field-line equation
in this system reads $d x / dz = \delta B_x / B_0$. Following the
established Kubo statistical formalism for random processes, the
field-line (FL) mean square displacement (MSD) can be written as
\be \left< \left( \Delta x (z) \right)^2 \right> = {1 \over B_0^2}
\int_{0}^{z} d z' \; \int_{0}^{z} d z'' \; R_{xx} (z',z'')
\label{s2eq2} \ee where we have employed $\Delta x = x(z) - x(0)$
and the $xx-$component of the magnetic correlation tensor $R_{xx}
(z',z'') = \left< \delta B_x (\vec{x} (z')) \delta B_x^{*}
(\vec{x} (z'')) \right>$; the real part of the right-hand side
(\emph{rhs}) will be understood. Assuming homogeneous turbulence
$R_{xx} (z',z'') = R_{xx} (|z' - z''|)$, one obtains \be
\left<\left( \Delta x (z) \right)^2 \right> = {2 \over B_0^2}
\int_{0}^{z} d z' \; (z - z') R_{xx} (z') \label{s2eq4} \ee with
$R_{xx} (z') = \left< \delta B_x (\vec{x} (z')) \delta B_x^{*}
(\vec{x} (0)) \right>$. Since the correlation tensor is itself
dependent of $\vec{x}(z) = (x(z),y(z),z)$ this is an implicitly
nonlinear integral equation for the magnetic field-line space
topology. In order to evaluate Eq. (\ref{s2eq4}), one has to
determine the correlation tensor $R_{xx}$. A Fourier
transformation leads to \be R_{xx} (z) = \int d^3 k \int d^3 k^{'}
\left< \delta B_x (\vec{k}) \delta B_x^{*} (\vec{k}^{'}) e^{i
\vec{k} \cdot \vec{x} (z) - i \vec{k}^{'} \cdot \vec{x} (0)}
\right> \, , \label{s3eq1} \ee or, adopting Corrsin's independence
hypothesis \cite{cor59} \bdm
R_{xx} (z) & = & \int d^3 k \; \int d^3 k^{'} \; \left< \delta B_x (\vec{k}) \delta B_x^{*} (\vec{k}^{'}) \right> \nonumber\\
& \times & \left< e^{i \vec{k} \cdot \vec{x} (z) - i \vec{k}^{'}
\cdot \vec{x} (0)} \right>  \, . \label{s3eq2} \edm Assuming that
the magnetic fields for different wave vectors are uncorrelated
$\left< \delta B_x (\vec{k}) \delta B_x^{*} (\vec{k}') \right> =
P_{xx} (\vec{k}) \delta (\vec{k} - \vec{k}')$ one is led to \be
R_{xx} (z) = \int d^3 k \; P_{xx} (\vec{k}) \left< e^{i \vec{k}
\cdot \Delta \vec{x} (z)} \right> \label{s3eq4} \ee with $P_{xx}
(\vec{k}) = < \delta B_x (\vec{k}) \delta B_x^{*} (\vec{k}) >$.
For the sake of analytical tractability, in order to evaluate the
characteristic function $< e^{i \vec{k} \cdot \Delta \vec{x} (z)}
>$, we shall assume Gaussian statistics for the field-lines, thus
\be \left< e^{i \vec{k} \cdot \Delta \vec{x} (z)} \right> =
e^{-{1\over 2} \left< \left( \Delta x (z) \right)^2 \right> k_x^2
-{1\over 2} \left< \left( \Delta y (z) \right)^2 \right> k_y^2
+ik_{\parallel} z}. \label{s4eq1} \ee For axisymmetric turbulence
$\left<(\Delta x)^2\right> = \left<(\Delta y)^2 \right>$, so Eq.
(\ref{s2eq4}) takes the form \bdm
& & \left< \left( \Delta x (z) \right)^2 \right> = {2 \over B_0^2} \int d^3 k \; P_{xx} (\vec{k}) \nonumber\\
& \times & \int_{0}^{z} d z' \; (z - z') \cos (k_{\parallel} z')
e^{-{1 \over 2} \left< \left( \Delta x (z') \right)^2 \right>
k_{\perp}^2}. \label{s2eq4ab} \edm Upon differentiation with
respect to $z$, we find for the field-line MSD \bdm
& & {d \over d z} \left< \left( \Delta x (z) \right)^2 \right> = {2 \over B_0^2} \int_{0}^{z} d z' \; R_{xx} (z') \nonumber\\
& = & {2 \over B_0^2} \int d^3 k \; P_{xx} (\vec{k}) \nonumber\\
& \times & \int_{0}^{z} d z' \; \cos (k_{\parallel} z')
e^{-{1\over 2} \left< \left( \Delta x (z') \right)^2 \right>
k_{\perp}^2}. \label{s2eq4abc} \edm A second differentiation leads
to \bdm
{d^2 \over d z^2} \left< \left( \Delta x (z) \right)^2 \right> & = & {2 \over B_0^2} R_{xx} (z) \nonumber\\
& = & {2 \over B_0^2} \int d^3 k \; P_{xx} (\vec{k}) \nonumber\\
& \times & \cos (k_{\parallel} z) e^{-{1 \over 2} \left< \left(
\Delta x (z) \right)^2 \right> k_{\perp}^2}. \label{s4eq2} \edm
This ODE was obtained, relying on no other assumptions than
Corrsin's hypothesis and Gaussian FL statistics. In combination
with different turbulence models, it provides a general basis for
the determination of the FL-MSD, thus allowing for a quantitative
description of field-line wandering. In the following, we shall
consider a specific example, as well as various limits of this
description.
{\bf
\section{Analytical and numerical results for slab/2D composite turbulence}
}
A two-component turbulence model has been proposed as a realistic
model for solar wind turbulence \cite{bie96}. Within this model,
the turbulent fields are described as a superposition of a slab
model ($\vec{k} \parallel \vec{B}_0$) and a two-dimensional (2D)
model ($\vec{k} \perp \vec{B}_0$). In the following, we shall
evaluate the field-line MSD, in view of identifying an comparing
among pure slab and composite slab/2D geometry.
{\bf
\subsection{Analytical results for pure slab geometry}
A first approach consists in assuming \emph{slab} turbulence
statistics. The $xx-$component of the correlation tensor then
reads \be P_{xx}^{slab} (\vec{k}) = g^{slab}(k_{\parallel})
{\delta (k_{\perp}) \over k_{\perp}} \, , \label{slabcorr} \ee
where we assume \be g^{slab} (k_{\parallel}) = {C(\nu) \over 2
\pi} l_{slab} \delta B_{slab}^2 (1 + k_{\parallel}^2
l_{slab}^2)^{-\nu} \label{slabwave} \ee for the slab wave
spectrum. Here $C(\nu)$ is a normalization constant given by \bdm
C(\nu) = {1 \over 2 \sqrt{\pi}} {\Gamma (\nu) \over \Gamma
(\nu-1/2)} \edm (see e.g. \cite{ShaKo07a}), $l_{slab}$ is the
slab-bendover-scale, $2\nu$ is the inertial-range spectral index,
and $\delta B_{slab}^2/B_0^2$ determines the relative strength of
slab turbulence. For the spectrum of Eq. (\ref{slabwave}), it is
straightforward to show that the slab result behaves diffusively,
as \be \left< \left( \Delta x (z) \right)^2 \right> = 2
\kappa_{slab} \, \mid z \mid \label{diffslab} \ee with the slab
field-line diffusion coefficient \be \kappa_{slab} = \pi C(\nu)
l_{slab} {\delta B_{slab}^2 \over B_0^2}. \label{kappaslab} \ee
This is an exact result, readily obtained upon analytical
evaluation of Eq. (\ref{s2eq4abc}) in the limit $\mid z \mid
\rightarrow \infty$.

\subsection{Analytical results for slab/2D composite geometry}
In a \emph{hybrid (composite) slab/2D} model, the correlation
tensor is assumed to have the form $P_{xx} (\vec{k}) =
P_{xx}^{slab} (\vec{k}) + P_{xx}^{2D} (\vec{k})$ (following the
definitions above). Here we used the slab correlation tensor of
Eq. (\ref{slabcorr}) and the 2D correlation tensor defined by \be
P_{xx}^{2D} (\vec{k}) = g^{2D} (k_{\perp}) {\delta (k_{\parallel})
\over k_{\perp}} \left( 1 -{k_x^2 \over k^2} \right) \, .
\label{s51eq4} \ee Eq. (\ref{s4eq2}) thus becomes \bdm
& & {d^2 \over d z^2} \left< \left( \Delta x \right)^2 \right> = {2 \over B_0^2} R_{xx}^{slab} (z) \nonumber\\
& + & {2 \pi \over B_0^2} \int_{0}^{\infty} d k_{\perp} \; g^{2D}
(k_{\perp}) e^{-{1 \over 2} \left< \left( \Delta x \right)^2
\right> k_{\perp}^2}. \label{s51eq5} \edm Focusing on high values
of the position variable $\mid z \mid \rightarrow \infty$, hence
$\left<( \Delta x)^2 \right> \rightarrow \infty$, the main
contribution to the integral comes from very low values of the
integration variable. Assuming a quasi-constant behavior of the 2D
spectrum in the energy-range we may approximate as \bdm
& & {d^2 \over d z^2} \left< \left( \Delta x \right)^2 \right>_{\mid z \mid \rightarrow \infty} \simeq {2 \over B_0^2} R_{xx}^{slab} (\mid z \mid \rightarrow \infty) \nonumber\\
& + & {2 \pi \over B_0^2} g^{2D} (0) \int_{0}^{\infty} d k_{\perp}
\; e^{-{1 \over 2} \left< \left( \Delta x \right)^2 \right>
k_{\perp}^2}. \label{s51eq6old} \edm It can easily be
demonstrated, by evaluating Eq. (\ref{s3eq1}) for pure slab
geometry, that the slab correlation function $R_{xx}^{slab}$ for
the spectrum of Eq. (\ref{slabwave}) has the asymptotic behaviour
\be R_{xx}^{slab} (\mid z \mid \rightarrow \infty) \sim \left(
{l_{slab} \over \mid z \mid} \right)^{1-\nu} e^{-\mid z \mid /
l_{slab}}. \label{slabcorrelation} \ee Obviously the first
contribution in Eq. (\ref{s51eq6old}) is much smaller than the
second, thus one obtains \bdm {d^2 \over d z^2} \left< \left(
\Delta x \right)^2 \right>_{\mid z \mid \rightarrow \infty} &
\simeq & {2 \pi \over B_0^2} \sqrt{\pi \over 2} {g^{2D} (0) \over
\sqrt{ \left< \left( \Delta x \right)^2 \right>}}. \label{s51eq6}
\edm It is straightforward to show that Eq. (\ref{s51eq6}) is
solved by \be \left< \left( \Delta x \right)^2 \right> \simeq
\left[ {9 \pi \over 2 B_0^2} \sqrt{\pi \over 2} g^{2D} (0)
\right]^{2/3} \mid z \mid^{4/3} \label{s51eq7} \ee in the limit
$\mid z \mid \rightarrow \infty$. To proceed we use a form for
$g^{2D} (k_{\perp})$ similar to the slab spectrum (see Eq.
(\ref{slabwave} )) \be g^{2D} (k_{\perp}) = {2 C(\nu) \over \pi}
l_{2D} \delta B_{2D}^2 \, (1 + k_{\perp}^2 l_{2D}^2)^{-\nu}.
\label{s51eq8} \ee Here $l_{2D}$ is the 2D-bendover-scale, $2\nu$
is the inertial-range spectral index, and $\delta B_{2D}^2/B_0^2$
determines the relative strength of 2D turbulence. Combining this
model spectrum with the above, we find \be \left< \left( \Delta x
\right)^2 \right> = \left[ 9 C(\nu) \sqrt{\pi \over 2} l_{2D}
{\delta B_{2D}^2 \over B_0^2} \right]^{2/3} \mid z \mid^{4/3}
\label{s51eq9} \ee which is clearly a \emph{non-diffusive} result.
As demonstrated, field-line wandering behaves superdiffusively if
the slab/2D composite model is employed for the turbulence
geometry. }
\subsection{Numerical evaluation}
Eq. (\ref{s2eq4ab}) is an integral-equation for the MSD of the
field-lines. By using Eqs. (\ref{s51eq4}) and (\ref{s51eq8}) for
$P_{xx}^{2D} (\vec{k})$, in addition to Eqs. (\ref{slabcorr}) and
(\ref{slabwave}) for $P_{xx}^{slab} (\vec{k})$, the total
correlation tensor $P_{xx} (\vec{k})$ for the composite model is
specified.

We have evaluated Eq. (\ref{s2eq4ab}) numerically, adopting a 20\%
slab-/80\% 2D- composite turbulence model. For the sake of
comparison, we have also computed the result in the pure-slab
turbulence case for $\delta B_{slab}^2/B_0^2=1$. For the same
cases of study, we have seen above that the analytical result for
two-component turbulence is given by Eq. (\ref{s51eq9}), whereas
the pure slab result is given by (\ref{diffslab}).

\begin{figure}[t]
\begin{center}
\epsfig{file=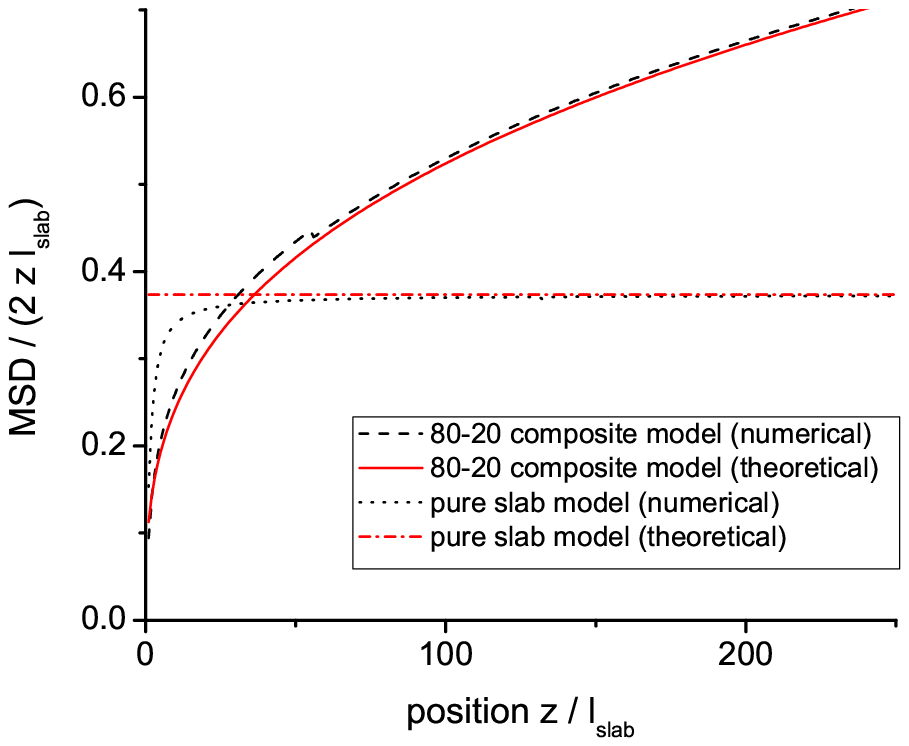, width=225pt}
\end{center}
\caption{(Color online) The running diffusion coefficients $\left<
\left( \Delta x (z) \right)^2 \right> / (2 z l_{slab})$ as a
function of $z/l_{slab}$ are depicted. The analytical result
(solid line) is compared to the numerical result (dashed line) for
20 \% slab / 80 \% 2D composite geometry. The analytical slab
result (dash-dotted line) and the numerical slab result (dotted
line) are also provided, for reference. The analytical and
numerical results are are obviously in good agreement.}
\label{figure1}
\end{figure}

The numerical results are compared with the analytical results
(presented above) in Fig. \ref{figure1}, by depicting the running
diffusion coefficients $< \left( \Delta x (z) \right)^2 > / (2 z
l_{slab})$ as a function of $z/l_{slab}$. An excellent agreement
is witnessed among the theoretical and the numerical results, in
both cases. The analytical expression (\ref{s51eq9}) derived above
for the asymptotic behavior of the field-line MSD thus appears to
be valid in the composite turbulence model also and, in fact,
bears a contributions which soon exceeds the diffusive result of
the slab-model significantly: cf. the upper curve(s) in Fig. 1 to
the lower, constant curve(s).
\section{Further results and limits}
We shall complete our study by considering various limiting cases,
in which previous theoretical results are recovered. Although
these results are not new, it is interesting \emph{per se} to
demonstrate that they can be obtained as appropriate limits from
Eqs. (\ref{s2eq4ab}-\ref{s4eq2}).
\subsection{The initial free streaming regime}
For small values of the position variable ($z \rightarrow 0$)
hence expecting $\left<(\Delta x)^2\right> \rightarrow 0$, we
obtain from Eq. (\ref{s4eq2}) \be {d^2 \over d z^2} \left< \left(
\Delta x \right)^2 \right> \approx {2 \over B_0^2} \int d^3 k \;
P_{xx} (\vec{k}) = 2 {\delta B_x^2 \over B_0^2}. \label{s5eq1} \ee
Assuming vanishing conditions for both the MSD and its derivative
at zero, one obtains \be \left< \left( \Delta x \right)^2 \right>
= {\delta B_x^2 \over B_0^2} z^2 \label{s5eq2} \ee so a strong
superdiffusive FL wandering regime is  clearly found for small
$z$, independent of the turbulence model adopted. {\bf For pure
slab geometry it can easily be demonstrated that the initial free
streaming solution is valid for $\mid z \mid \ll l_{slab}$. This
means that Eq. (\ref{s5eq2}) is valid for length scales which
correspond to the inertial range of the turbulence wave spectrum.}
\subsection{Quasilinear theory for field-line random walk}
Within QLT, one may replace the MSD on the right hand side of Eq.
(\ref{s4eq2}) by the unperturbed field-lines (thus taking $\left<
(\Delta x )^2 \right> = 0$), which yields \be {d^2 \over d z^2}
\left< \left( \Delta x \right)^2 \right> = {2 \over B_0^2} \int
d^3 k \; P_{xx} (\vec{k}) \cos (k_{\parallel} z). \label{s51eq3}
\ee QLT is thus apparently only exact for pure-slab turbulence,
where $P_{lm} \sim \delta (k_{\perp})$; cf. Eq. (\ref{slabcorr}).
On the other hand, within QLT one finds for pure 2D and slab/2D
composite turbulence [adopting Eqs. (\ref{s51eq4}) and
(\ref{s51eq3})] that $\left< (\Delta x )^2 \right> = (\delta B_x /
B_0)^2 z^2$, in disagreement with the nonlinear result obtained
above. Thus, QLT fails to describe field-line wandering in the
two-component model. Presumably, quasilinear theory might thus
also not apply in other non-slab models.
\subsection{The diffusion limit}
Combining Eq. (\ref{s2eq4abc}) with the assumption of diffusion
for the magnetic field-lines (i.e. explicitly setting $\left< (
\Delta x )^2 \right> = 2 \kappa \mid z \mid$) one obtains \be
\kappa = {1 \over B_0^2} \int d^3 k \; P_{xx} (\vec{k})
\int_{0}^{\infty} d z \; \cos (k_{\parallel} z) e^{- \kappa z
k_{\perp}^2}. \label{matt1} \ee The $z-$integral can easily be
solved to give \be \kappa = {1 \over B_0^2} \int d^3 k \; P_{xx}
(\vec{k}) {\kappa k_{\perp}^2 \over k_{\parallel}^2 + (\kappa
k_{\perp}^2)^2}. \label{difflimit} \ee This formula is correct if
field-line random walk behaves diffusively and if the small length
scales of the initial free streaming regime are unimportant.

For two-component turbulence Eq. (\ref{difflimit}) can easily be
evaluated; we find \be \kappa = \kappa_{slab} + {\kappa_{2D}^2
\over \kappa} \, , \label{matt2} \ee using the slab diffusion
coefficient $\kappa_{slab}$ of Eq. (\ref{kappaslab}) and \be
\kappa_{2D}^2 = {1 \over B_0^2} \int d^3 k \; k_{\perp}^{-2}
P_{xx}^{2D} (\vec{k}) \, . \label{matt3} \ee Eq. (\ref{matt2}) is
a quadratic equation in $\kappa$, which may easily be solved to
get \be \kappa = {\kappa_{slab} + \sqrt{\kappa_{slab}^2 + 4
\kappa_{2D}^2} \over 2} \, . \label{matt4} \ee We note that this
coincides with the result derived earlier by Matthaeus \textit{et
al.} \cite{mat95}.  For pure slab turbulence, we have
$\kappa_{2D}=0$, so the expected limit $\kappa=\kappa_{slab}$ is
recovered. On the other hand, for pure 2D geometry, i.e. for
$\kappa_{slab}=0$, one finds $\kappa=\kappa_{2D}$. Thus the
parameter $\kappa_{2D}$ can be identified with the
\emph{``diffusion coefficient''} for pure 2D turbulence. {\bf
However, adopting the standard spectrum of Eq. (\ref{s51eq8}), one
clearly obtains a diverging result: \bdm
\kappa_{2D}^2 & = & 2 C(\nu) l_{2D} {\delta B_{2D}^2 \over B_0^2} \int_{0}^{\infty} d k_{\perp} k_{\perp}^{-2} \left( 1 + l_{2D}^2 k_{\perp}^2 \right)^{-\nu} \nonumber\\
& \rightarrow & \infty. \label{divergent} \edm The only
possibility to prevent this singularity is the introduction of a
finite box-size $L_{2D} \gg l_{2D}$ of the 2D fluctuations. Then
we find \bdm \kappa_{2D}^2 & = & 2 C(\nu) l_{2D} {\delta B_{2D}^2
\over B_0^2}
\int_{L_{2D}^{-1}}^{\infty} d k_{\perp} k_{\perp}^{-2} \left( 1 + l_{2D}^2 k_{\perp}^2 \right)^{-\nu} \nonumber\\
& = & 2 C(\nu) {\delta B_{2D}^2 \over B_0^2} \left( {L_{2D} \over
l_{2D}} \right)^{1+2\nu} {1 \over 2 \nu + 1} \, \, l_{2D}^2
\nonumber\\
& \times & \;_2 F_1 \left( \nu, \nu + {1 \over 2}, \nu + {3 \over
2}; - {L^2 \over l_{2D}^2} \right) \, ,  \edm where the integral
was solved by applying exact relations which can be found in Ref.
\cite{Grad}. Because of $L_{2D} \gg l_{2D}$ we consider the
asymptotic limit of the hypergeometric function $_2 F_1 (a,b; x)$
(see \cite{stegun}) to find for $\kappa_{2D}$ the expression \bdm
\kappa_{2D} = \sqrt{2 C(\nu) \over 2 \nu + 1} \sqrt{l_{2D} L_{2D}}
{\delta B_{2D} \over B_0}. \label{matt5} \edm Within the diffusion
theory for FLRW, the diffusion coefficient is strongly controlled
by the box-size $L_{2D}$. In the light of the results presented in
section III., the true reason for the divergent behavior (see Eq.
(\ref{divergent})) may be the \emph{superdiffusive} nature of
field-line random walk. It appears that Eqs. (\ref{matt4}) and
(\ref{matt5}) fail to provide an appropriate description of the
composite turbulence, in contrast with the generalized nonlinear
theory presented in this paper. }
{\bf
\section{Is a superdiffusive behaviour of FLRW reasonable?}
A key result of the current article is the superdiffusive
behaviour of FLRW for the slab/2D composite turbulence model (see
Eq. (\ref{s51eq9}) and Fig. 1). However, this nondiffusive result
is clearly in disagreement with the assumption of diffusive field
line wandering employed in several previous articles (see e.g.
\cite{mat95}).

The superdiffusive result derived here is based on two {\it ad
hoc} assumptions which cannot be deduced from first principles:
\begin{enumerate}
\item A random phase (or Corrsin type) approximation was applied
(see Eq. (\ref{s3eq2})). \item A Gaussian distribution of field
lines was assumed (see Eq. (\ref{s4eq2})).
\end{enumerate}
Eq. (\ref{s51eq9}) therefore provides an accurate result,
suggesting a superdiffusive behavior of FLRW in real systems,
provided that the latter two assumptions hold. However, one might
draw the conclusion that the non-diffusivity deduced in this
article is a consequence of (two) inaccurate approximations. We
shall attempt to remove this ambiguity in the discussion that
follows.

In another article (\cite{ShaKo07b}) we show that already for pure
slab fluctuations where FLRW can be described without any
assumptions or approximations, a superdiffusive result can be
obtained be replacing the standard spectrum of Eq.
(\ref{slabwave}) by a decreasing spectrum in the energy range.
According to these results, superdiffusion indeed arises as the
{\it normal}, by default behavior of FLRW. Classical (Markovian)
diffusion can only be obtained in very special limits (wave
spectrum exactly constant in the energy range and pure slab
geometry).

The slab correlation function decays rapidly with increasing
distance (see Eq. (\ref{slabcorrelation})). For decreasing wave
spectra in the energy range and for non-slab models, however, we
find a much weaker (non-exponential) decrease of the magnetic
correlation functions. This weaker descrease of magnetic
correlation functions provides a physical explanation of the
nondiffusivity discovered in the current article.

A useful application but also a test of the superdiffusive result
is cosmic ray scattering perpendicular to the mean magnetic field.
In \cite{ShaKo07a} the superdiffusive result obtained here is
combined with a compound transport model based on the assumption
that the guiding centers of the charged cosmic rays follow the
magnetic field lines. In this article a quite good agreement
between test particle simulations and the combination of compound
transport and superdiffusion of FLRW was found. This agreement can
also be seen as a proof of the validity of the two assumption
(Corrsin's hypothesis, Gaussian statistics) used in this article
to deduce the superdiffusive result.

Furthermore, it is argued in \cite{ShaKo07a} that for a diffusive
behaviour of field line wandering one always obtains a strong
subdiffusive behaviour of cosmic ray perpendicular scattering
which cannot be true. Thus, the superdiffusive result deduced in
the current article is not only a result obtained under certain
assumptions. It seems that the real field lines indeed behave
superdiffusively and that this superdiffusivity is essential for
understanding charged particle propation.

From a purely fundamental point of view, the
statistical-mechanical description of FLRW in magnetized plasmas
bears certain generic physical characteristics which are
encountered in various physical contexts. Our analytical findings
are qualitatively reminiscent of earlier results by Isichenko on
advection-diffusion problems \cite{Isio92}. Physically speaking,
diffusive behavior of turbulence implies a fast decay of
Lagrangian correlations, sufficiently fast for the limit $<
(\Delta x )^2 >/2t$ to converge. In fluid (non-plasma)
environments, long scale random velocity  fluctuations (depending
on boundary conditions) may dominate the motion of a fluid element
and may thus cause the limit to diverge. On  the other hand,
turbulent magnetized plasma systems are characterized by a
plethora of nonlinear mechanisms, intertwined with the complex FL
topology, which may add up in such a way that anomalous
(non-classical) diffusion become the rule in plasmas. Persisting
correlations among successive FL displacements are thus built-up,
resulting in the superdiffusive behavior exposed in our work. This
seems to suggest a general tendency of turbulent magnetized
systems, so that previous diffusive FLRW considerations (obtained
under various assumptions) should rather be treated with some
caution.

It may be added that a superdiffusive result for FLRW was also
obtained in various previous works, where distinct methodologies
were employed. In Ref. \cite{Jo73}, the FL separation across the
field was studied by adopting an ad hoc form for the magnetic
correlation tensor; its behavior was found to be very sensitive to
the exact form of the power spectrum, while a parabolic
(ballistic) regime was shown to rule for steep spectra. In Ref.
\cite{Skill74}, geometric (non-statistical) arguments were
employed to study FLRW in galactic cosmic rays. A superdiffusive
behavior was also found, independently from the turbulence
spectrum (though mainly shallow spectra were considered therein).
Similar findings were reported in Ref. \cite{Nar01}, in the
context of thermal conduction in galactic clusters. A chaotic
magnetic FL behavior (explicitly assumed to obey a Lyapunov
behavior over a wide range of length scales), i.e. assuming an
exponential FL divergence, was there shown to enhance
perpendicular thermal conduction significantly, essentially
increasing thermal conductivity in the transverse direction
(previously thought to be much weaker than the parallel one) by a
factor (up to) 5, as compared to the Spitzer value \cite{Spitzer}.
Still in the galactic thermal conduction context, Refs.
\cite{Chan04, Maron04} have adopted a double diffusion concept,
where particle diffusion along the field was related to FLRW in
space, in order to study thermal conduction and electron diffusion
along the magnetic field; a combined phenomenological,
Fokker-Planck and numerical approach, FL displacement was shown to
increase fast over length scales which are a multiple of the
characteristic turbulence scale. Also note Ref. \cite{Matt03} (and
the exhaustive discussion therein) for a nonlinear treatment of
the perpendicular diffusion of charged particles (not FLRW).

 }

\section{Conclusion}
A generalized nonlinear formulation has been presented, to
describe field-line wandering (random walk) in general
magnetostatic systems consisting of a statistical (or turbulent)
component $\delta B$ and a nonstatistical uniform component $B_0$.
Assuming vanishing parallel component of the turbulent field
($\delta B_z =0$), applying the Corrsin approximation, and
assuming Gaussian statistics, a general ODE was deduced for the
field-line MSD (Eq. (\ref{s4eq2})) in axisymmetric and homogeneous
turbulent plasmas.

Adopting the two-component turbulence model which was suggested by
Bieber et al. (\cite{bie96}) as a realistic model for solar wind
turbulence, it was demonstrated systematically that FLRW behaves
superdiffusively. Specifically, a weakly \emph{superdiffusive}
behavior of the mean square deviation was found in the form $<
(\Delta x)^2 > \sim \mid z \mid^{4/3}$. This result was compared
to previous results obtained via different assumptions, namely the
quasilinear result ($< (\Delta x)^2 > \sim z^{2}$) and the
diffusive result $< (\Delta x)^2 > \sim \mid z \mid$ of Ref.
\cite{mat95}.

Extending the considerations in this article, the generalized
nonlinear formulation presented here may be applied in a
description of non-axisymmetric turbulence and/or other (e.g.,
anisotropic) turbulence geometries.

Since field-line random walk is of major importance in the
description transport of charged particles in astrophysical
plasmas, these results are significant in cosmic ray space physics
and astrophysics. Concluding, the analytical expression(s) derived
in this article provide a useful toolbox, which extends existing
transport theoretical models and might be essential in inspiring
forthcoming ones.
\begin{acknowledgments}
This research was supported by Deutsche Forschungsgemeinschaft
(DFG) under the Emmy-Noether Programme (grant SH 93/3-1). As a
member of the {\it Junges Kolleg} A. Shalchi also aknowledges
support by the Nordrhein-Westf\"alische Akademie der
Wissenschaften.
\end{acknowledgments}
{}
\newpage

\end{document}